# INFERENCE OF WATER WAVES SURFACE ELEVATION FROM HORIZONTAL VELOCITY COMPONENTS USING PHYSICS INFORMED NEURAL NETWORKS (PINN)

[1]OMAR SALLAM, [2]MIRJAM FÜRTH

[1,2]Texas A&M University, 727 Ross St, College Station, TX 77843
E-mail.[1]osallam@tamu.edu,[2]furth@tamu.edu

**Abstract** - In this paper, a mathematical model is presented to infer the wave free surface elevation from the horizontalvelocity components using Physics Informed Neural Network (PINN). PINN is a deep learning framework to solve forward and inverse Ordinary/Partial Differential Equations (ODEs/PDEs). The model is verified by measuring a numerically generated Kelvin waves downstream of a KRISO Container Ship (KCS). The KCS Kelvin waves are generated using two phase Volume of Fluid (VoF) Computational Fluid Dynamics (CFD) simulation with OpenFOAM. In addition, the paperpresented the use of the Fourier Features decomposition of the Neural Network inputs to avoid the spectral bias phenomena; Spectral bias is the tendency of Neural Network to converge towards the low frequency solution faster than the high frequency one. Fourier Features decomposition layer showed an improvement for the model learning, as the model was able to learn the high and low frequency components simultaneously.

**Keywords** - Two phase flow, Volume of Fluid, Physics Informed Neural Networks, OpenFOAM.

## I. INTRODUCTION

Designing marine vehicles, offshore platforms or costal protection structures requires a precise knowledge of the sea-state in the operating location (Chakrabarti, 1994). Sea-state data are often described in statistical manner such as significant waveheight $H_s$, significant wave period $T_s$ and directional energy spectrum $E(f, \kappa, \theta)$, where f is the wave frequency, $\kappa$ is the wave number and $\theta$ is the wave direction.

The statistical data of the sea-state are achieved using in-situ wave measurement techniques or remote sensing wave measurement techniques;in-situ techniques include wave buoys, wave gaugesor pressure transducers,whereas remote sensing techniques include radar, lidar and stereovision systems(Holthuijsen, 2010).Remote sensing techniques often provide more informative sea-state data compared to in-situ techniques, as the free surface elevation measurement is implemented on scanned areas compared to in-situ techniques that collects data at localized points;hence the remote sensing techniquesmeasurement includes statistical data for the spatial/temporal domainscompared to the in-situ techniques that are limited to temporal domain analysis at certain localized points. Measurement of sea-state around the world is dominated by in-situ techniques due to the high cost and complexity of operation and implementation of remote sensing techniques.

In this paper, we aim to presenta mathematical model for a newly introduced remote sensing techniqueof measuring the random oceanic waves surface. The introduced concept is based on inferring the oceanic random surface elevation from measured horizontalvelocity components using Physics Informed Neural Network (PINN)(Raissi, Perdikaris, & Karniadakis, Physics informed deep learning (part i): Data-driven solutions of nonlinear partial differential equations, 2017). This concept assumes that the horizontalvelocity components can be attained using a pre-calibrated camera(Sallam & Fürth, 2023).

### 1.1. Physics Informed Neural Network (PINN)

Physics Informed Neural Network (PINN) was first introduced by Raissi et.al(Raissi, Perdikaris, & Karniadakis, Physics informed deep learning (part i): Data-driven solutions of nonlinear partial differential equations, 2017)as a deep learning tool for solving forward or inverse problems formulated in Ordinary/Partial Differential Equations (ODEs/PDEs). For forward problems, the Initial/Boundary Conditions (ICs/BCs) are well defined while the solution in the domain $\Omega$ is unknown. On the other hand, for inverse problems, the ICs/BCs are not well defined, but sparsely solution data are known for the spatial or temporal domains $(T, \Omega)$(Raissi, Deep hidden physics models: Deep learning of nonlinear partial differential equations, 2018).Fig.1 shows the architecture of the PINN where $(\mathbf{x} \in \Omega, t \in T)$ are the independent variables inputs and $\mathbf{u}$ is the dependent variable to be predicted. The residuals of the set of governing ODEs/PDEs is $N(\mathbf{u}(\mathbf{x}, t))$, where N is any general differential operator. The partial derivatives of the ODEs/PDEs are evaluated by the automatic differentiation (Baydin, Pearlmutter, Radul, & Siskind, 2018) available in machine learning platforms such as PyTorch(Paszke, et al., 2019) or TensorFlow (Abadi, et al., 2016). The optimization algorithms in the Neural Network (NN) task is tuning the NN's hyperparameter $\theta$ to minimize residuals of the ODEs/PDEs to be solved and minimize the error between the sparsely known data $u_{sampled}$ and the network predictions $u_{net}$.The NN's hyperparameter $\theta$ include weights w and biases b.

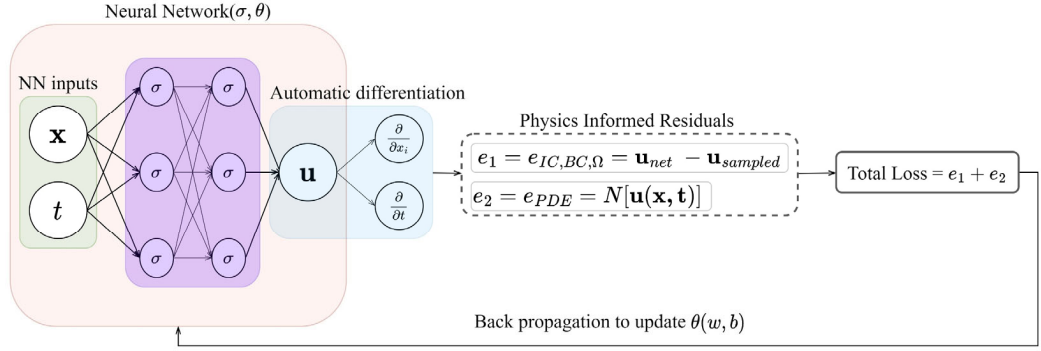

**Fig.1.Physics Informed Neural Network (PINN)architecture for solving a general ODE/PDE**

### 1.2. Fourier Features-Neural Networks (FF-NN)

For Neural Networks (NNs), solving problems that exhibit spatial or temporal multiple frequency scales, spectral bias phenomenon takes place (Rahaman, et al., 2019). Spectral bias is the tendency of the NNs to converge towards the low frequency solution component faster than the higher frequency ones, and convergence to the high frequency solutions is not guaranteed even with higher training iterations (Tancik, et al., 2020). Tancik et.al(Tancik, et al., 2020), have proposed a solution to address the spectral bias problem in NNs called Fourier Features-Neural Network (FF-NNs). In FF-NNs, the input data are transformed or mapped to a vector of superposed periodic functions such as sine and cosine, see Fig.3. This transformation layer is known as the Fourier Feature embeddings, and the frequency vector elements of the Fourier Feature embedding layer are sampled from a normal distribution. The authors have effectively utilized this Fourier Feature embedding approach to improve the reconstruction of high-frequency-colored images in neural networks, enabling the reconstruction of high-frequency features such as hair and fur,while the conventional NNs failed to do even with large iteration numbers. Building on FF-NN,Wang et al (Wang, Wang, & Perdikaris, 2021)proposed the Multi Scale-Spatio Temporal-Fourier Features- Physic Informed Neural Network(MS-ST-FF-PINN) architecture to tackle the spectral bias phenomenon in ODEs/PDEs that exhibit multi frequencies component scales in spatial or temporal domains. In MS-ST-FF-PINN, both spatial and temporal domains inputs ($\mathbf{x}, t$) are fed to a Fourier decomposition layerto modulate on higher frequency signals, the frequencies of these signals are drawn fromnormal distributions with standard deviation $\sigma_{f_x}$ for the spatial input and $\sigma_{f_t}$ for thetemporal input.The proposedMS-ST-FF-PINN architecture has been tested on Poisson, wave and Gray Scott equations and successfully learned the multi scale frequency solution components of the mentionedPDEs(Wang, Wang, & Perdikaris, 2021). For time independent problems the architecture has no spatio/temporal terms and abbreviated as FF-PINN.

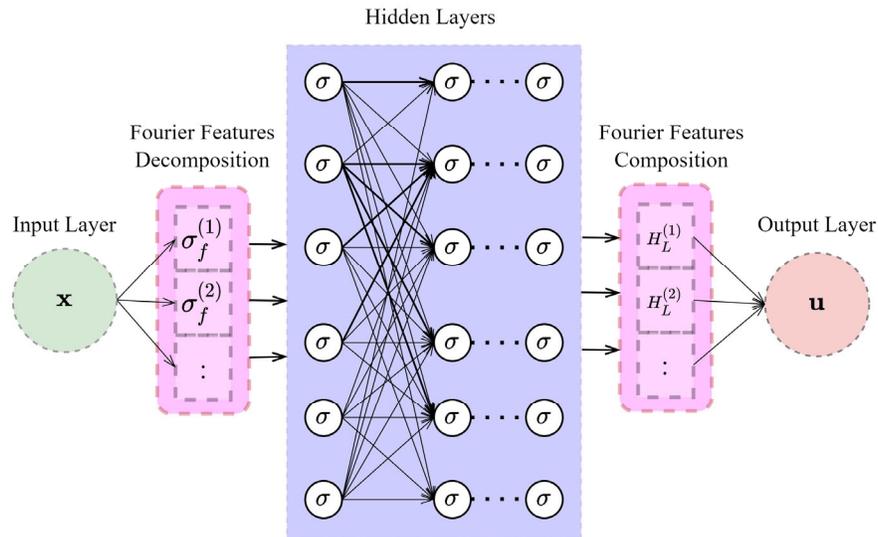

**Fig.2.Fourier Features-Neural Network (FF-NN) architecture, the Fourier Features embedding is to overcome the spectral bias phenomenon.**

## II. METHODOLOGY

In this section, the water wave elevation model is illustrated based on the PINN and FF-PINN architectures. In addition, the modelnumerical verification test case setup is presented, where a numerical waveis generated using OpenFOAM(Jasak, Jemcov, Tukovic, & others, 2007).

### 2.1 Physics Informed Neural Network (PINN) model for free surface water waves inference.

For any general PINN problem, three main sets must be defined. model inputs, model outputs and the ODEs/PDEs to be satisfied; for inverse PINN problems, additional domain training sampled data must be fed to the network.

Using PINN to infer the wave elevation$\eta$ from the free surface horizontalvelocity components ($u_1, u_2$) is an inverse PINN problem, the free surface horizontalvelocity components represent the sampled training sampled data while the 3 velocity components and the wave elevation represent the PINN output. The PDEs set to be satisfied consists of a modified version of the Momentum equation and the Kinematic Free Surface boundary conditions (KFS) as shown in equations (1) and (2)respectively; the derivation of these equations is presented in (Salmon, 2008) where the problem dimension is reduced from 3D to 2D.

Fig.3 and Fig.4show the PINN and the MS-ST-FF-PINN architectures for the 2D free surface problem. It is obvious that the two architectures have the same inputs, outputs and set of PDEs, however the MS-ST-FF-PINN has extra two layers (Spatio temporal Fourier Features decompositionlayer and spatio temporal Fourier Features Composition layer); in the first one, the network spatial and temporal inputs are modulated to higher frequency signals separately, while in the second one a superposition of the modulated signals is implemented to predict the network outputs in the spatio/temporal domains.

$$\frac{\partial u_i}{\partial t} + u_j \frac{\partial u_i}{\partial x_j} = g_3 \frac{\partial \eta}{\partial x_i} \; ; \; i \& j \in \{1,2\} \tag{1}$$

$$u_3 = \frac{\partial \eta}{\partial t} + u_j \frac{\partial \eta}{\partial x_j} ; j \in \{1,2\} \tag{2}$$

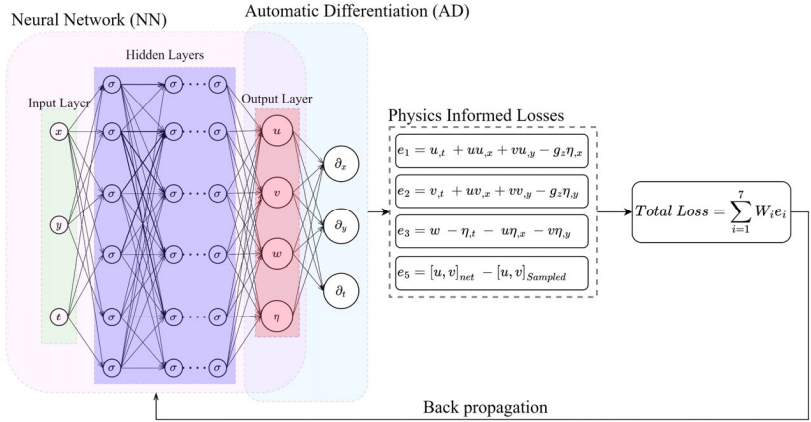

**Fig.3. PINN architecture for 2D wave free surface model.**

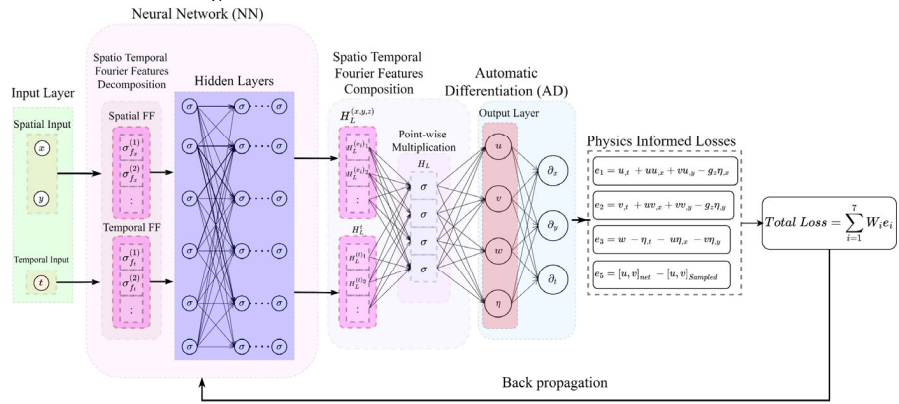

**Fig.4.MS-ST-FF-PINN architecture for 2D wave free surface model. In this architecture, two extra layers are added (Fourier Feature decomposition and Fourier Feature composition) layers.**

## 2.2 Numerical generation of Kelvin wave pattern of KCS model using OpenFOAM.

To test and verify the PINN and FF-PINN architectures for inferring the wave elevation η from the horizontalvelocity components $(u_1, u_2)$; the two architectures are tested on a numerically generated Kelvin wave pattern downstream of a KRISO Container Ship (KCS)(Kim, Van, & Kim, 2001) at 0.26 Froude's number (Fr). KCS model parameters is illustrated inTable.1. The simulation is implemented using Interfoam(Deshpande, Anumolu, & Trujillo, 2012); Interfoam is a two-phase Volume of Fluid (VoF) solver for incompressible and isothermal fluids in OpenFOAM(Jasak, Jemcov, Tukovic, & others, 2007).

The free surface is simulated using the Volume of Fluid method with a single fluid mixture assumption. This method solves for the volume phase fraction α transport equation (5) of the primary fluid, which in this case is water. Additionally, the continuity and momentum equations shown in equations (3) and (4) are also solved. The InterFoam library utilizes the PIMPLE algorithm for velocity pressure coupling(Holzmann, 2016). The density ρ and dynamic viscosity μ of the mixture fluid are determined based on the mixture phase fraction, as shown in equations (6) and (7). The subscript a or w represents air or water, respectively. The surface tension force is denoted by $f_{\sigma i}$.

$$\frac{\partial u_j}{\partial x_j} = 0 \quad (3)$$

$$\frac{\partial \rho u_i}{\partial t} + \frac{\partial (\rho u_i u_j)}{\partial x_j} = -\frac{\partial p}{\partial x_i} + \mu \frac{\partial^2 u_i}{\partial x_j \partial x_j} + \rho g_i + f_{\sigma i} \quad (4)$$

$$\frac{\partial \alpha}{\partial t} + \frac{\partial (u_j \alpha)}{\partial x_j} = 0 \quad (5)$$

$$\rho = \alpha \rho_w + (1-\alpha)\rho_a \quad (6)$$

$$\mu = \alpha \mu_w + (1-\alpha)\mu_a \quad (7)$$

| | |
|---|---|
| Scale ratio | 1/31.6 |
| Speed [m/sec] | 2.1964 |
| Length Over All (LOA)[m] | 2.72 |
| Length Water Line (LWL)[m] | 2.28 |
| Breadth [m] | 1.0190 |
| Depth [m] | 0.6013 |
| Draft [m] | 0.3418 |
| Froude number (Fr) | 0.26 |

**Table.1 KCS model parameters.**

Due to symmetry, only half of the domain is discretized and simulated. The computational domain background grid is generated using the blockMesh utility in OpenFOAM with dimensions ($39 \times 15 \times 6$ m$^3$) and resolution ($120 \times 60 \times 4$) cells. SnappyHexMesh meshing utility (Gisen, 2014)is used to construct the mesh around the KCS model and apply refinement at the free surface region resulting in $9.5 \times 10^6$ finite volume cells, seeFig.5. In Fig.5, the red cells represent water ($\alpha = 1.0$) and blue cells represent air $\alpha = 0.0$.

The inlet flow velocity $U_\infty$ is 2.1964 m/sand the vessel Froude number is $Fr = \frac{U_\infty}{\sqrt{gL}} = 0.26$, where g is the gravitational acceleration and L is the KCS length at water line (LWL = 7.28 m). The problem is simulated for 20 seconds, the initial time step is 0.001 seconds, this time step is automatically adjustable to satisfy the maximum set Courant number =1.0.

Fig.6shows the Kelvin wave pattern generated downstream the KCS model after 20 seconds of simulation time. The pink shaded rectangle area($10 \times 14$ m$^2$)is the study region and starts 5 m downstream of the KCS model where the horizontalvelocities $(u_1, u_2)$ are sampled to train the both PINN and FF-PINN architectures.

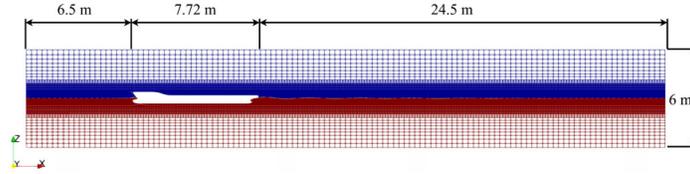

Fig.5.Numerical wave tank length and height, grid is clustered at the free surface region, red cells represent water and blue cells represent air.

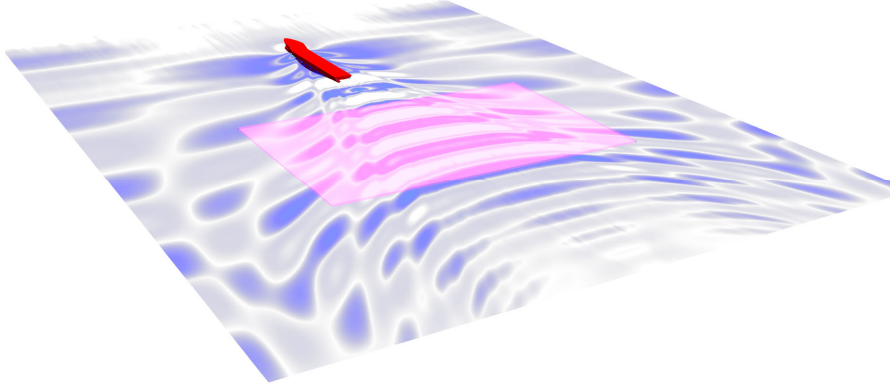

Fig.6. kelvin wave pattern downs stream of the KCS model at 0.26 Froude number.

## III. MODEL VERIFICATION AND RESULTS

To test and verify the proposed model, a single snapshot of the of the OpenFOAM simulation results is used for the model training on the free surface horizontalvelocity components at t = 20 sec as shown in Fig.6. Training on a single snapshot makes the problem time independent and the temporal derivative term $\frac{\partial(.)}{\partial t}$ vanishes in equations (1) and (2).Using the Python libraryDeepXDE(Lu, Meng, Mao, & Karniadakis, 2021), both PINN and FF-PINN architectures are implemented to train the embedded NNs on the horizontalvelocity components $(u_1, u_2)$ and infer the wave elevation η and the vertical velocity component $u_3$.Both PINN and FF-PINN training were implemented on a Nvidia Quadro RTX 4000 GPU with 8GB dedicated memory. For both architectures, Adams algorithm (Kingma & Ba, 2014)is used as an optimizer to minimize the total loss function represented by the summation of the PDEs residuals and the mean square error between the OpenFOAMhorizontalvelocity components $(u_1, u_2)$ and the predicted velocity components by the NNs$(u_1, u_2)_{net}$. The optimizer learning rate is $10^{-3}$ and the NN is trained for $4 \times 10^4$ iterations. $16 \times 10^3$ random points are sampled from the OpenFOAM solution for the horizontal velocity components training, while $2 \times 10^4$ collocation points are uniformlysampled in the domain to satisfy the PDEs (1) and (2).In the FF-PINN architectures, the Fourier Features decomposition layer modulates the spatial inputs $(x_1, x_2)$ to higher frequencies signals, these higher frequencies are sampled from two normal distributions with standard deviations $\sigma_{f_x} \in \{0.1, 1.0\}$.

After both architectures reached the maximum training iterations = $2 \times 10^4$, the output variables$(u_1, u_2, u_3, \eta)$ were predicted at the same spatial coordinates of the OpenFOAM mesh for the free surface, this enables to compute a pointwise error between the OpenFOAM and the predicted solutions for the whole 2D study domain.

Fig.7 shows the OpenFOAM solution, PINN and the FF-PINN predictionsfor the problem output variables $(u_1, u_2, u_3, \eta)$. For the two free surface horizontal velocity components $(u_1, u_2)$ that both architectures have been trained on, PINN architecture was able to predict only the low frequency solution of the water disturbance due to the KCS model motion andfailed to learn the higher frequency solution of horizontal velocities due to the Kelvin wave pattern due to the spectral bias problem. On the other hand, FF-PINN architecture was able to learn both low/high frequency solutions for the horizontal velocities. PINN architecture completely failed to infer the vertical velocity component $u_3$or the surface elevation η and its predictionshows a smooth low frequency behavior for the two variables.FF-PINN architecture was able to infer the vertical velocity component $u_3$ and the surface elevation η of the Kelvin wave patternin a good agreement with the OpenFOAM solution, however FF-PINN imposed additional wiggles(high frequency signals)to the $u_3, \eta$ predictions that made the prediction not as smooth as the OpenFOAM solution.

Fig.8shows the training loss (absolute relative error) for the horizontal velocity components $(u_1, u_2)$, the plot shows that the FF-PINN has lower error compared to the PINN architecture. Similarly, Fig.9 and Fig.10 shows the mean relative absolute error of the predictions/inference of PINN and FF-PINN architectures for the vertical velocity component $u_3$ and the surface elevation $\eta$ at different training cycles, the Fourier Features embeddings showed a dramatic improvement for the accuracy of predictions.

The PDEs residuals are shown in Fig.11; surprisingly, PINN architecture showed lower residuals for the problem set of PDEs although it was not able to learn the high frequency component of the horizontal velocity components $(u_1, u_2)$ and totally failed to infer the vertical velocity component $u_3$ and the surface elevation $\eta$. This means that PINN architecture converged to a solution that satisfied the set of PDEs (trivial solution) but completely different from the correct solution, this one of the drawbacks of PINN where no unique solution exists(Leiteritz & Pflüger, 2021).

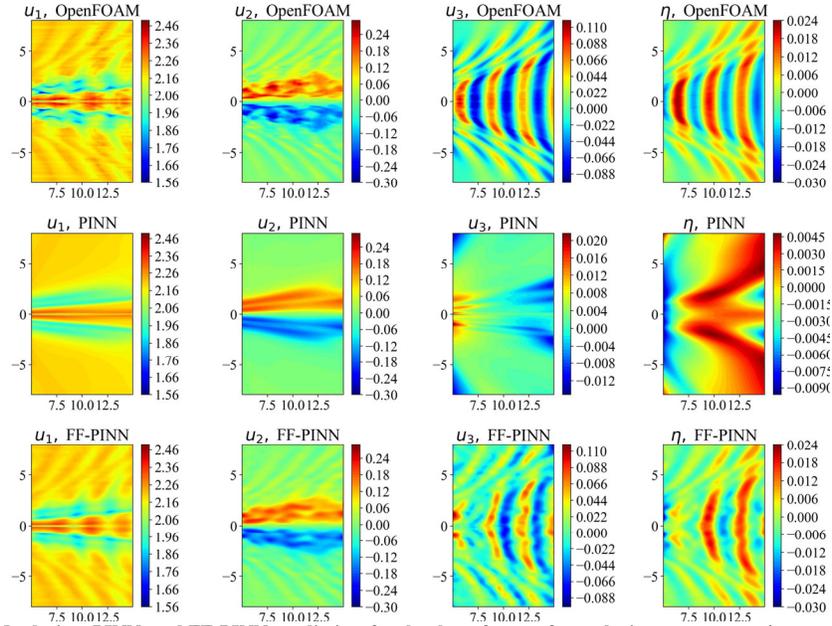

**Fig.7.OpenFOAM solution, PINN, and FF-PINN predictions for the three free surface velocity components $(u_1, u_2, u_3)$ and the wave elevation $\eta$. FF-PINN architecture was able to learn both high/low frequency solutions compared to the PINN architecture that converged to low frequency solutions.**

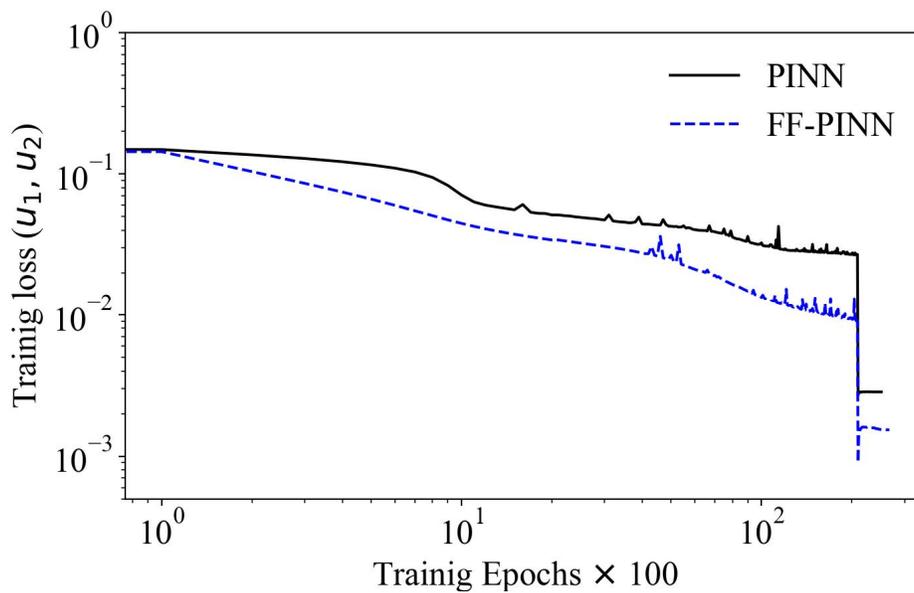

**Fig.8. Summation of training loss for the water free surface horizontal velocity components $(u_1, u_2)$. FF-PINN architecture showed better learning capability throughout the training iterations compared to the PINN architecture.**

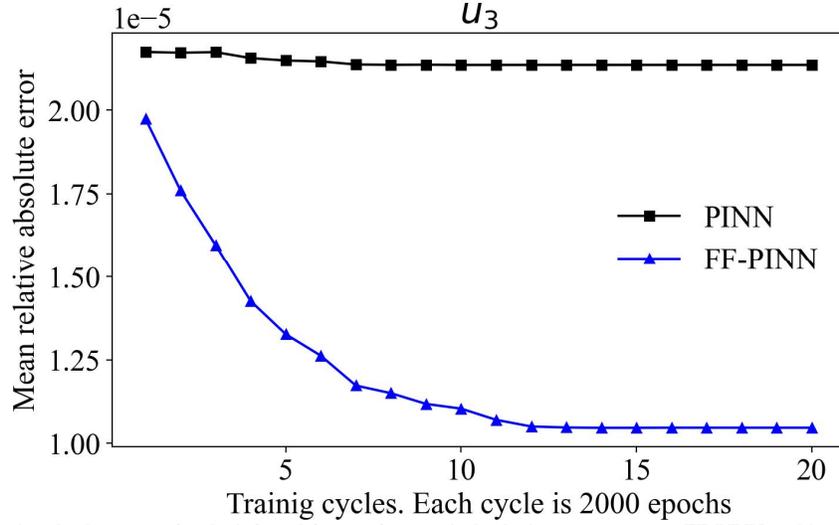

**Fig.9.** Mean relative absolute error for the inferred free surface vertical velocity component $u_3$. FF-PINN architecture shows lower error compared to the PINN.

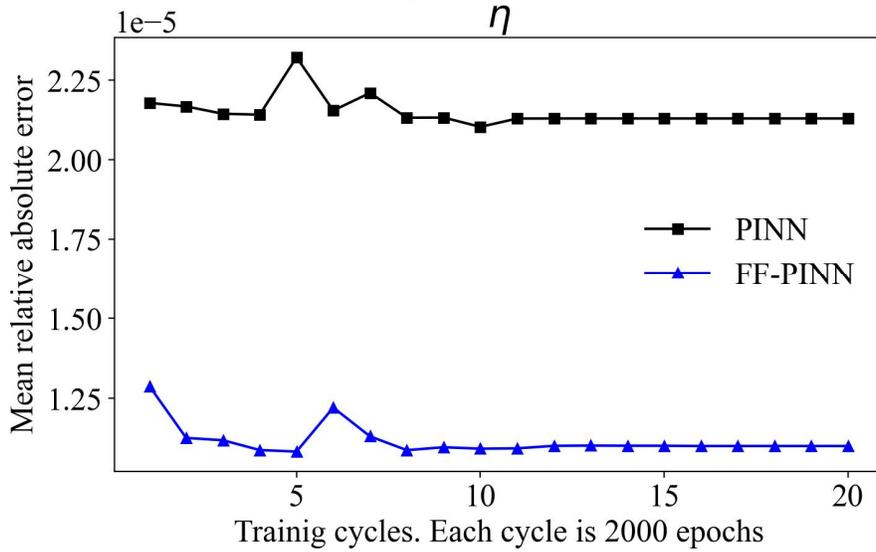

**Fig.10.** Mean relative absolute error for the inferred $\eta$. FF-PINN architecture shows lower error compared to the PINN.

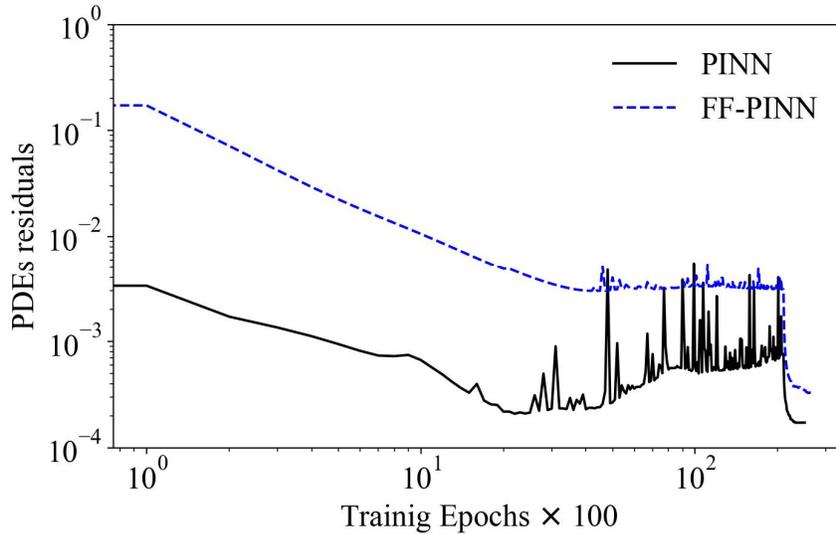

**Fig.11.** PDEs residuals for PINN and FF-PINN architectures.

## IV. CONCLUSION

In this paper, a Physics Informed Neural Network (PINN) model is presented to infer the water wave surface elevation and the vertical wave's velocity component from the measured two horizontal velocity components of the water free surface. The model is tested on a numerically generated Kelvin wavesdownstream of a KRISO Container Ship (KCS) at 0.26 Froude number. The numerical wave generation is achieved by two phase Volume of Fluid (VoF) simulation using InterFoam solver in OpenFOAM. The paper also studied the effect of the Fourier Features embedding to the PINN architecture (FF-PINN). The results showed that the PINN architecture with no Fourier Features embeddings struggled to learn the high frequency components in the horizontal velocity components of the water free surface due to the spectral bias phenomenon and totally failed to infer the wave elevation or the free surface vertical velocity component. On the other hand, embedding the Fourier Features to the PINN architecture (FF-PINN) improved the learning capability and controlled the spectral bias phenomenon. FF-PINN was able to learn the high/low frequency component solutions simultaneously of the horizontal free surface velocity components. In addition, FF-PINN architecture was able to infer the free surface vertical velocity component and the wave elevation in good agreement with the OpenFOAMsolution. In future work, this mathematical model will be experimentally tested on regular and irregular generated waves. The model will be validated by a stereo vision based depth camera system that scan the water wave free surface in the study domain.

★★★